\documentstyle[amssymb,aps,psfig,epsfig,graphics,twocolumn]{revtex}

\begin{document}
\title{Superchemistry: dynamics of coupled atomic and molecular Bose-Einstein condensates}
\author{D. J. Heinzen, R. H. Wynar}
\address{{\it Department of Physics, University of Texas, Austin, }Texas, 78712}
\author{P. D. Drummond, and K. V. Kheruntsyan}
\address{{\it Department of Physics, University of Queensland, }\\
St. Lucia, {\it Queensland 4067, Australia}}
\date{March 17, 2000}
\maketitle

\begin{abstract}
We analyze the dynamics of a dilute, trapped Bose-condensed atomic gas
coupled to a diatomic molecular Bose gas by coherent Raman transitions. This
system is shown to result in a new type of `superchemistry', in which giant
collective oscillations between the atomic and molecular gas can occur. The
phenomenon is caused by stimulated emission of bosonic atoms or molecules
into their condensate phases.\newline
\end{abstract}

The experimental observation of dilute gas Bose-Einstein condensation (BEC)
is revolutionizing low-temperature physics \cite{BEC}. BEC represents the
ultimate limit for the cooling of a gas, since nearly all of the atoms may
occupy the condensate ground state. Rather than being identifiable as single
particles, these atoms coherently populate a matter wave field, and can be
coupled out of the condensate to produce an `atom laser' \cite{Atom laser}.
This suggests the possibility of nonlinear atom optics. Due to atomic
interactions, even a single-species condensate may exhibit nonlinear wave
behavior analogous to self-phase modulation in optics \cite{Deng}. Here, we
propose that a more general type of nonlinearity may occur through
atom-molecule coupling, and show that this coupling may result in the
formation of a molecular Bose condensate through stimulated emission of
molecular bosons.

More generally, we define `superchemistry' as {\it the coherent stimulation
of chemical reactions via macroscopic occupation of a quantum state by a
bosonic chemical species}. In other words, `superchemistry' results in
greatly enhanced, non-Arrhenius chemical kinetics at ultra-low temperatures.
In the simplest case of $A+B\rightarrow C$ reactions, there are three
possibilities for the quantum statistics of the components: $bb\rightarrow b$%
, $bf\rightarrow f$, and $ff\rightarrow b$, where $b$ stands for bosonic and 
$f$ for fermionic. In all three cases stimulated emission can occur.
Interestingly, the latter two of these cases correspond to well-known
quantum-field theories, the Lee-Van Hove model of meson theory, and the
Friedberg-Lee model of high-T$_{C}$ superconductivity \cite{Lee et al.}. In
this Letter we consider a chemical system of the first type where bosonic
enhancement of the chemical dynamics is the strongest. We develop the theory
of coherently interacting atomic and molecular condensates needed to
describe this process, and consider a specific coupling mechanism based on
stimulated free-bound Raman transitions \cite{Raman-Molecules-BEC}.

We begin with the usual quantum field theory Hamiltonian for a
noninteracting (atomic or molecular) species ($i$), in a well-defined
internal state: 
\begin{eqnarray}
\hat{H}^{(0)} &=&\sum_{i}\int d^{3}{\bf x}\left[ \frac{\hbar ^{2}}{2m_{i}}%
|\nabla \hat{\Phi}_{i}({\bf x})|^{2}\right.   \nonumber \\
&&\left. +\,\left( V_{i}({\bf x})+E_{i}\right) \hat{\Phi}_{i}^{\dag }({\bf x}%
)\hat{\Phi}_{i}^{{}}({\bf x})\right] \,.  \label{H_0}
\end{eqnarray}
Here, $m_{i}$ is the mass, $V_{i}({\bf x})$ is the trapping potential, and $%
E_{i}$ the internal energy of species $i$. The particles also interact
through collisions. We consider particle number-conserving collisions
mediated by an inter-species potential $U_{ij}({\bf x})$, and non-conserving
collisions mediated by an effective potential $\chi _{ijk}({\bf x})$. The
first of these nonlinear terms describes the well-known intra-species
repulsion or attraction, as well as inter-species couplings \cite
{Inter-species}. It is desirable to introduce a momentum cutoff to simplify
the field theory \cite{Abrikosov-Dzyaloshinski}, and to replace $U_{ij}({\bf %
x})$ by an effective pseudopotential $U_{ij}\delta ({\bf x})$. This
describes low-energy $S$-wave scattering only. Similarly, the potential $%
\chi $ can be replaced by an equivalent $S$-wave pseudopotential, again with
the proviso that a momentum cutoff is introduced at the level of $k_{m}\sim
a^{-1}$, where $a$ is the longest scattering length in the problem. The
result is an effective quantum field theory \cite{BECSolitons} in which: 
\begin{eqnarray}
\hat{H}_{eff}^{(c)} &=&\frac{1}{2}\sum_{ij}\int d^{3}{\bf x}\left[ \hat{\Phi}%
_{i}^{\dag }({\bf x})\hat{\Phi}_{j}^{\dagger }({\bf x})U_{ij}\hat{\Phi}%
_{j}^{{}}({\bf x})\hat{\Phi}_{i}^{{}}({\bf x})\right] ,  \nonumber \\
\hat{H}_{eff}^{(nc)} &=&\frac{1}{2}\sum_{ijk}\int d^{3}{\bf x}\left[ \hat{%
\Phi}_{i}^{\dag }({\bf x})\hat{\Phi}_{j}^{\dagger }({\bf x})\chi _{ijk}\hat{%
\Phi}_{k}^{{}}({\bf x})+H.c.\right] .  \label{H_nc eff}
\end{eqnarray}
In the diagonal case, $U_{ii}=4\pi \hbar ^{2}a_{i}/m_{i}$, where $a_{i}$ is
the $i$-th species scattering length. In the present work, we assume that
the trap potential $V_{i}({\bf x})$ is harmonic: $V_{i}({\bf x}%
)=(m_{i}/2)\omega _{i}^{2}|{\bf x}|^{2}$, where $\omega _{i}$ represents the
rotationally symmetric trap-oscillation frequency for the $i$-th species.

The new feature introduced here is the particle number non-conserving
potential $\chi$. Terms like this occur in nonlinear optics, where they
describe parametric processes of sub- and second-harmonic generation, which
change the photon number \cite{Butcher}. While matter is clearly not created
or destroyed in low temperature experiments, an analogous effect can occur
where two atoms are converted into one molecule. Inside a Bose condensate,
this chemical conversion is dominated by coherent stimulated emission, in
which transitions are enhanced by the number of molecules already occupying
the ground state. This is completely different from the usual chemical
kinetics, which predicts that the rates of chemical reactions do not depend
on the number of particles in the product mode, and go to zero at low
temperatures according to the Arrhenius law. This type of classical
(Boltzmann) kinetic theory is inapplicable in BECs, where the particle
wavelength exceeds the interparticle spacing.

In general, the conversion process $i+j\rightleftarrows k$ will be
non-resonant. The exception, for which $E_{k}=E_{i}+E_{j}$, corresponds to a
Feshbach resonance; such resonances have recently been studied
experimentally \cite{Feshbach-experiment}. Alternatively, energy conserving
transitions are possible if $\chi_{ijk}$ has a harmonic time dependence. In
this paper, we consider the specific case of stimulated Raman coupling
induced by two laser fields {\bf E}$_{L1}$ and {\bf E}$_{L2}$ of frequencies 
$\omega_{L1}$ and $\omega_{L2}$, as illustrated in Fig. 1. This coupling
becomes resonant when the Raman detuning $\delta=(2E_{1}-E_{2})/\hbar-(%
\omega_{L2}-\omega _{L1})$ goes to zero. This allows coupling to a single
molecular state, which can be selected by the Raman laser frequencies.

The implication of these new terms is seen most easily by considering the
corresponding mean field equations, in which the operators are replaced by
their mean values, and a factorization is assumed. Elsewhere \cite
{BECSolitons}, using a variational technique, we have shown that this gives
a good estimate of the ground-state energy at high density - relative to the
spatially uncorrelated behavior that can occur at low densities
(corresponding to a Bose gas of dressed dimers). In the present case the
relevant equations are obtained, in a rotating frame, for the simplest case
of one atomic species $\phi _{1}$, together with a corresponding molecular
dimer species $\phi _{2}$: 
\begin{eqnarray}
i\hbar \dot{\phi}_{1} &=&-\frac{\hbar ^{2}}{2m_{1}}\nabla ^{2}\phi
_{1}+V_{1}({\bf x})\phi _{1}+U_{11}|\phi _{1}|^{2}\phi _{1}  \nonumber \\
&&+\;U_{12}|\phi _{2}|^{2}\phi _{1}+\chi \phi _{1}^{\ast }\phi _{2}^{{}},
\label{Eq-phi1} \\
i\hbar \dot{\phi}_{2} &=&-\frac{\hbar ^{2}}{2m_{2}}\nabla ^{2}\phi
_{2}+V_{2}({\bf x})\phi _{2}-\hbar \delta \phi _{2}+U_{22}|\phi
_{2}|^{2}\phi _{2}  \nonumber \\
&&+~U_{21}|\phi _{1}|^{2}\phi _{2}+\frac{1}{2}\chi ^{\ast }\phi _{1}^{2}.\ 
\label{Eq-phi2}
\end{eqnarray}
Here we assume that $U_{12}=U_{21}$ is the only number-conserving scattering
process, while $\chi \equiv \chi _{112}$ describes conversion of atoms into
molecules by stimulated Raman transitions. Many interesting dynamical
properties of these types of equations - including nonlinear oscillations,
non-equilibrium phase transitions, and soliton formation - have been
explored in nonlinear optics \cite{Malomed}. A novel feature here is the
presence of the trap potential which localizes the interaction volume.

We derive the Raman coupling coefficient $\chi$ for a simplified model of
the two-body interaction, in which the atoms interact in their electronic
ground state through a single Born-Oppenheimer potential $V_{g}(R)$.
Molecules are formed in a single bound vibrational state of energy $E_{2}$
with radial wave function $u_{2}(R)$. Two free atoms with zero relative
kinetic energy have a total energy $2E_{1}$, and a relative radial
wavefunction $u_{1}(R)$, normalized so that asymptotically $u_{1}\sim\sqrt{%
4\pi}(R-a_{1})$. We assume that the two laser fields ${\bf E}_{Li}=$ ${\bf E}%
_{0i}\cos(\omega _{Li}t)$ $(i=1,2)$ couple the ground electronic state to a
single electronically excited state described by a potential $V_{e}(R)$,
with Rabi frequencies $\Omega_{i}=|{\bf d}_{M}\cdot{\bf E}_{0i}|/\hbar$
where ${\bf d}_{M}$ is the electric dipole matrix element connecting these
two states. The excited state has vibrational levels $|v\rangle$ with energy 
$E_{v}$ and radial wave functions $u_{e,v}(R)$. The excited levels decay by
spontaneous emission at a rate $\gamma_{M}$. All bound levels are normalized
so that $\int dR|u_{e,v(2)}|^{2}=1$.

To proceed further, we first notice that the effective Hamiltonian in
first-order perturbation theory should reproduce the known behavior of two
atoms in a relative $S$-wave scattering process in the presence of an
external radiation field \cite{Fedichev etc.}. Here we recall that the
effective field theory has a momentum cutoff (otherwise the perturbation
theory would diverge for higher order terms). From these requirements we
obtain 
\begin{eqnarray}
\frac{U_{11}}{\hbar } &=&\frac{U_{0}}{\hbar }-\sum_{v}\left[ \frac{(\Omega
_{1})^{2}}{4\Delta _{v}}+\frac{(\Omega _{2})^{2}}{4\Delta _{v}^{(1)}}\right]
|I_{1,v}|^{2},  \label{U} \\
\frac{\chi }{\hbar } &=&-\frac{\Omega _{1}\Omega _{2}}{2\sqrt{2}}\sum_{v}%
\frac{I_{1,v}I_{2,v}^{\ast }}{\Delta _{v}}.  \label{chi}
\end{eqnarray}
Here $U_{0}/\hbar =4\pi \hbar a_{1}/m_{1}$, $\Delta
_{v}=(E_{v}-2E_{1})/\hbar -\omega _{L1}$, $\Delta
_{v}^{(1)}=(E_{v}-2E_{1})/\hbar -\omega _{L2}$, and $I_{j,v}$ are the
overlap integrals $I_{j,v}=\int dR\,u_{e,v}(R)u_{j}^{\ast }(R)$. In
addition, the molecular spontaneous emission leads to the incoherent
production of molecules in different states, together with atomic excited
state decays. Treating these as loss processes, we obtain additional terms
of form: 
\begin{eqnarray}
\dot{\phi}_{1} &=&-\alpha \phi _{1}+i\beta _{1}\phi _{1}-\Gamma _{1}|\phi
_{1}|^{2}\phi _{1},  \nonumber \\
\dot{\phi}_{2} &=&-\Gamma _{2}\phi _{2}+i\beta _{2}\phi _{2},
\label{decay eqs}
\end{eqnarray}
where the induced decay rates are: 
\begin{eqnarray}
\Gamma _{j} &=&\frac{\gamma _{M}}{8}\sum_{v}\left[ \frac{(\Omega _{j})^{2}}{%
\Delta _{v}^{2}}+\frac{(\Omega _{3-j})^{2}}{(\Delta _{v}^{(j)}{})^{2}}\right]
|I_{j,v}|^{2}\,\;\;\;\;\;\;(j=1,2),  \label{Gamma} \\
\alpha  &=&\frac{\gamma _{A}}{8}\sum_{i=1,2}\frac{(\Omega _{i}^{A})^{2}}{%
D_{i}^{2}}\;,  \label{alpha}
\end{eqnarray}
and $i\beta _{j}\phi _{j}$ is a light shift term, with 
\begin{eqnarray}
\beta _{1} &=&\sum_{i=1,2}\frac{(\Omega _{i}^{A})^{2}}{4D_{i}}\,,
\label{beta} \\
\beta _{2} &=&\sum_{v}\left[ \frac{(\Omega _{2})^{2}}{4\Delta _{v}}+\frac{%
(\Omega _{1})^{2}}{4\Delta _{v}^{(2)}}\right] |I_{2,v}|^{2}\,.
\end{eqnarray}
Here we have introduced $\Delta _{v}^{(2)}=(E_{v}-E_{2})/\hbar -\omega _{L1}$%
, $D_{i}=\omega _{0}-\omega _{Li}$, where $\omega _{0}$ is the resonance
frequency of the atomic transition between the dissociation limits of the
excited and ground potentials. Also, $\Omega _{j}^{A}=|{\bf d}_{A}\cdot {\bf %
E}_{0i}|/\hbar $ is the atomic Rabi frequency, ${\bf d}_{A}$ is the
transition dipole moment, and $\gamma _{A}$ is the atomic excited state
population decay rate. The Raman detuning at trap center for an atomic BEC
is $\tilde{\delta}\equiv \delta +\beta _{2}-2\beta _{1}+2(U_{11}/\hbar
)|\phi _{1}(0,0)|^{2}$.

Rotationally or vibrationally inelastic atom-molecule collisions may also
give rise to losses. The magnitude of these rates is presently unknown, and
we neglect them here. We note that these rates should decrease rapidly with
increasing molecular binding energy and go to zero in the molecular ground
state, so that it should be possible to obtain a very low rate by selection
of an appropriate level.

We have calculated $U_{11}$, $\chi$, $\Gamma_{i}$, $\beta_{i}$, and $\alpha$
for a $V_{g}(R)$ which closely approximates the $^{87}$Rb$_{2}$ ground $%
^{3}\Sigma_{u}^{+}$ potential, and a $V_{e}(R)$ which closely approximates
the $^{87}$Rb$_{2}$ 0$_{g}^{-}$ symmetry potential that connects to the 5$%
^{2}$S$_{1/2}$ + 5$^{2}$P$_{3/2}$ dissociation limit. Free-bound Raman
coupling of similar states in $^{85}$Rb$_{2}$ has previously been explored
experimentally \cite{spectroscopy}. In this calculation, $|2\rangle$ is the
bound state of $V_{g}$ corresponding to the vibrational quantum number $v=29$%
, with a binding energy of $-160.7$ GHz with respect to $2E_{1}$, and we
take $\gamma_{A}=3.7\times10^{7}s^{-1}$, $\gamma_{M}=2\gamma_{A}$, $\Omega
_{1}=2\times10^{10}$ s$^{-1}$, $\Omega_{2}=6.324\times10^{9}$ s$^{-1}$, and $%
\Omega_{i}^{A}=\Omega_{i}/\sqrt{2}$. We also choose $\omega_{L1}=\omega
_{0}-15.485$ cm$^{-1}$, where $\omega_{0}$ is the resonance frequency of the
5$^{2}$S$_{1/2}\leftrightarrow5^{2}$P$_{3/2}$ atomic transition. For these
parameters, we find that $\chi/\hbar\simeq7.6\times10^{-7}{\rm \,}$m$^{3/2}$%
/s, $\Gamma_{1}\simeq1.629\times10^{-23}\,$m$^{3} $/s, $\Gamma
_{2}\simeq304.4\,$s$^{-1}$, $\beta_{1}\simeq2.108\times10^{7}\,$s$^{-1}$, $%
\beta_{2}\simeq3.344\times10^{6}\,$s$^{-1}$, and $\alpha=134.06$ s$^{-1}$.
For typical, realizable BEC densities of $n\sim4\times10^{20}$ m$^{-3}$, we
find that $\chi n^{1/2}\gg\Gamma_{2},\Gamma_{1}n$ , as required for coherent
dynamics.

Based on these results, we have carried out simulations of Eqs. (\ref
{Eq-phi1}) and (\ref{Eq-phi2}), using $a=5.4\,$nm \cite{scattering length}, $%
U_{11}/\hbar\simeq4.96\times10^{-17}\,$m$^{3}$/s, and with additional terms
given by Eqs. (\ref{decay eqs}). The atom-molecule ($U_{12}$) and
molecule-molecule ($U_{22}$) scattering rates are neglected. Provided they
are not too strong, the effect of these terms will simply be to change the
condensate self-energies, and hence to modify the optimum Raman detuning. We
also choose $\omega_{1}/2\pi=\omega_{2}/2\pi=100\,$Hz, and an initial number
of atoms $N=5\times10^{5}$. We assume an initial condition of a pure atomic
BEC, with no molecules present, as given by the steady state of the standard
Gross-Pitaevskii equation in a trap.

The results are shown in Fig. 2. We observe giant oscillations between
atomic and molecular condensates, which take place on short time scales. The
integrated atomic and molecular numbers, shown in Fig. 3 (a), do not show
complete atom-molecule conversion, because the oscillation frequency is
higher in the center of the trap, due to the inhomogeneous atomic density.
The total number shows a small decay due to spontaneous emission.

The pronounced oscillation between an atomic and a molecular condensate
provides clear evidence of a long-range coherence effect. In contrast,
stimulated Raman photoassociation in a thermal cloud of atoms would not
produce similar collective oscillations, because the phases associated with
the individual atom/molecule conversion processes are random in a thermal
cloud. The effect is also very different from stimulated transitions between
two spin states of Bose-condensed atoms. \cite{spin-1/2 BEC}. That
conversion is linear in the atomic amplitudes, and therefore the stimulated
transition rate is equal to the single particle rate and independent of the
density \cite{nonlinear}. In contrast, stimulated atom-molecule conversion
in a condensate is nonlinear in the atomic and molecular amplitudes, and the
conversion rate scales with density as $\dot{n}_{i}\propto\chi n_{1}\sqrt{%
n_{2}}$.

Fig. 3 (b) shows the result of the calculation for a reduced atom number and
two times smaller density, but with the same effective initial Raman
detuning. The density-dependence of the superchemistry transitions is
evident in the comparison of Figs. 3(a) and 3(b), which shows that the
higher density cloud oscillates faster. This effect could be studied by
optically imaging the atomic cloud at a succession of times. Observation of
an essentially pure atomic condensate during one of the atom number maxima
would imply that the molecular gas which exists at a prior time must be
coherent. Important signatures of the coherent stimulation are a reaction
rate that initially increases with time (see insert, Fig. 3 (a)), and the
density dependence of the nonlinear oscillation period.

In summary, number-nonconserving interactions between bosonic atomic and
molecular condensates at ultra-low temperatures is predicted to result in a
form of `superchemistry' -- in which Bose-enhanced coherent quantum dynamics
replaces the usual chemical kinetics, giving a completely new type of
behavior. We find that giant collective oscillations can occur between
atomic and molecular Bose condensates. The phenomenon is the matter wave
analog to optical frequency doubling and parametric down conversion.
Interesting quantum behavior may also result, ranging from squeezed-state
generation \cite{Squeeze} to quantum soliton formation \cite{BECSolitons},
or even quantum chaos. Quantum effects may change short-distance
correlations, and need to be included in any treatment going beyond the mean
field theory. In the future, the highly specific nature of these proposed
Bose-enhanced reactions could open the way for new types of
quantum-controlled chemical synthesis, or other novel and unexpected
quantum-phase dependent phenomena.

D.J.H. and R.H.W. gratefully acknowledge support of the R. A. Welch
Foundation, the NASA Microgravity Research Division, and the NSF. P. D. and
K. K. acknowledge support from the Australian Research Council, and the NSF
(Grant PHY94-07194).

\bigskip

\begin{figure}[h]
\vspace{-0.8 cm}
\centerline{\hbox{
\epsfig{figure=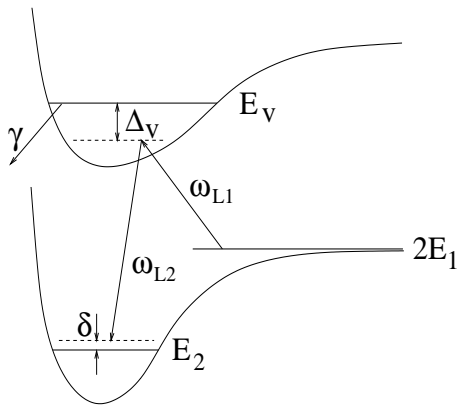,height=5.3 cm,width=6 cm}}}
\vspace{0.3 cm}
\caption{Schematic diagram of the Raman photoassociation.}
\end{figure}

\bigskip

\begin{figure}[h]
\vspace{-0.8 cm}
\centerline{\hbox{
\epsfig{figure=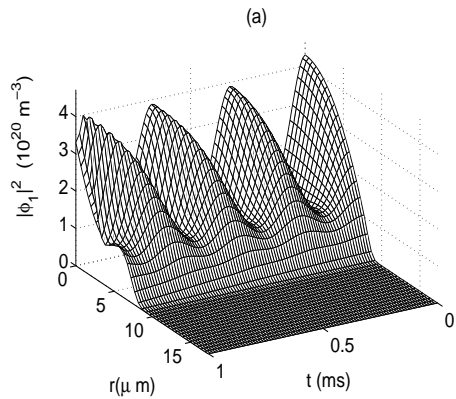,height=5.3 cm,width=6 cm}}}
\centerline{\hbox{
\epsfig{figure=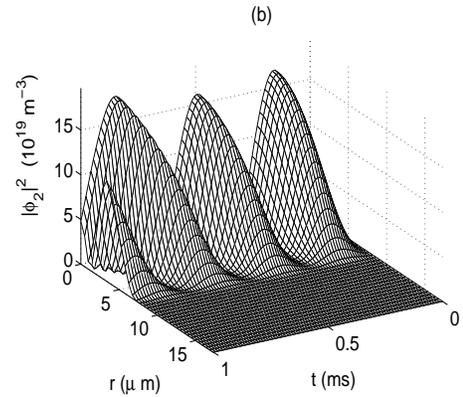,height=5.3 cm,width=6 cm}}}
\vspace{0.3 cm}
\caption{ Densities $|\phi _{i}({\bf x},t)|^{2}$ for the atomic (a) and
molecular (b) species as depending on time and radial distance $r=|{\bf x}|$
from the trap center, for $\delta =3.879\times 10^{7}\,$s$^{-1}$ (so that $%
\delta +\beta _{2}-2\beta _{1}=-2.8\times 10^{4}\,$s$^{-1}$).}
\end{figure}

\bigskip

\begin{figure}[h]
\vspace{-0.8 cm}
\centerline{\hbox{
\epsfig{figure=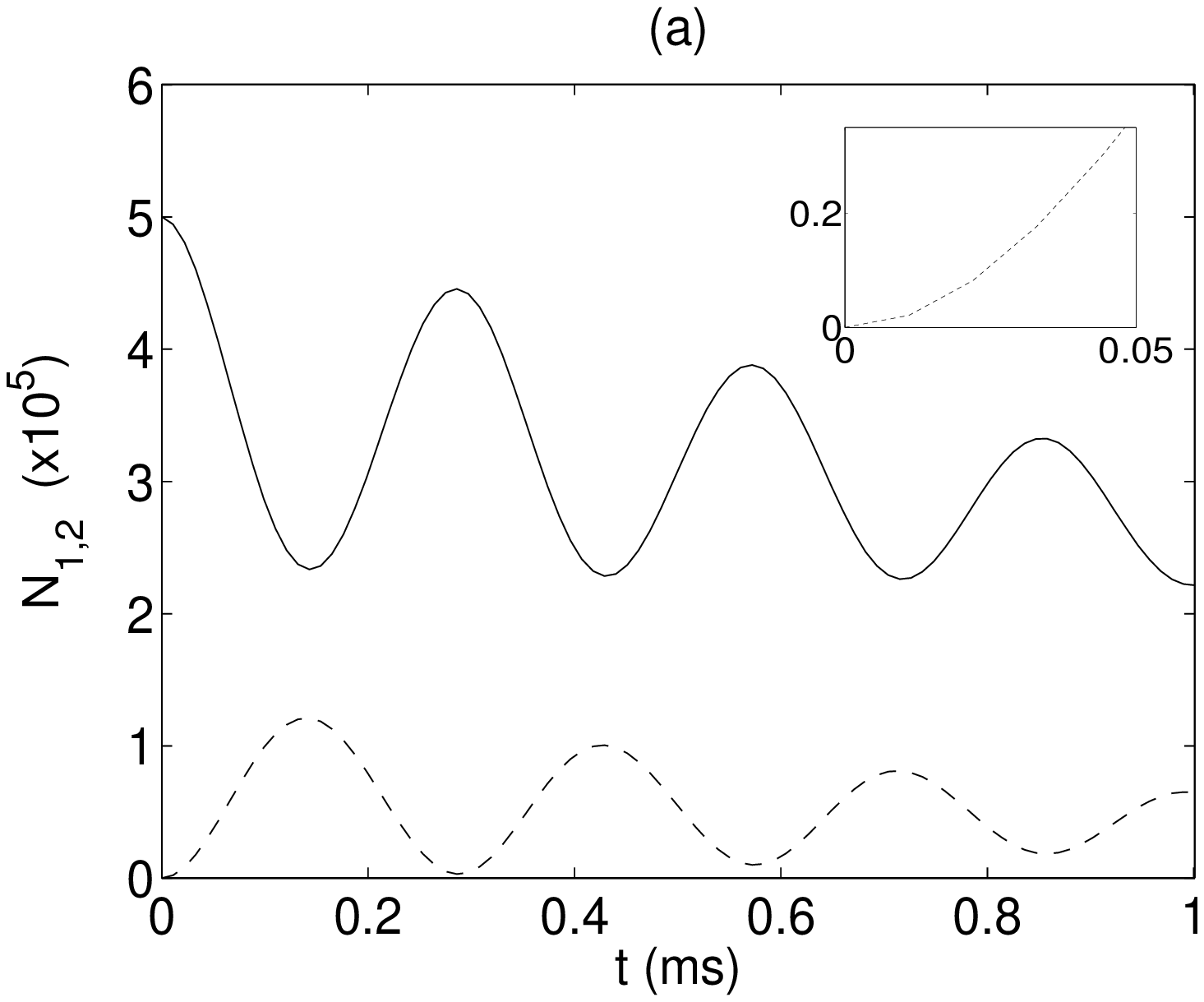,height=5 cm,width=5.7 cm}}}
\centerline{\hbox{
\epsfig{figure=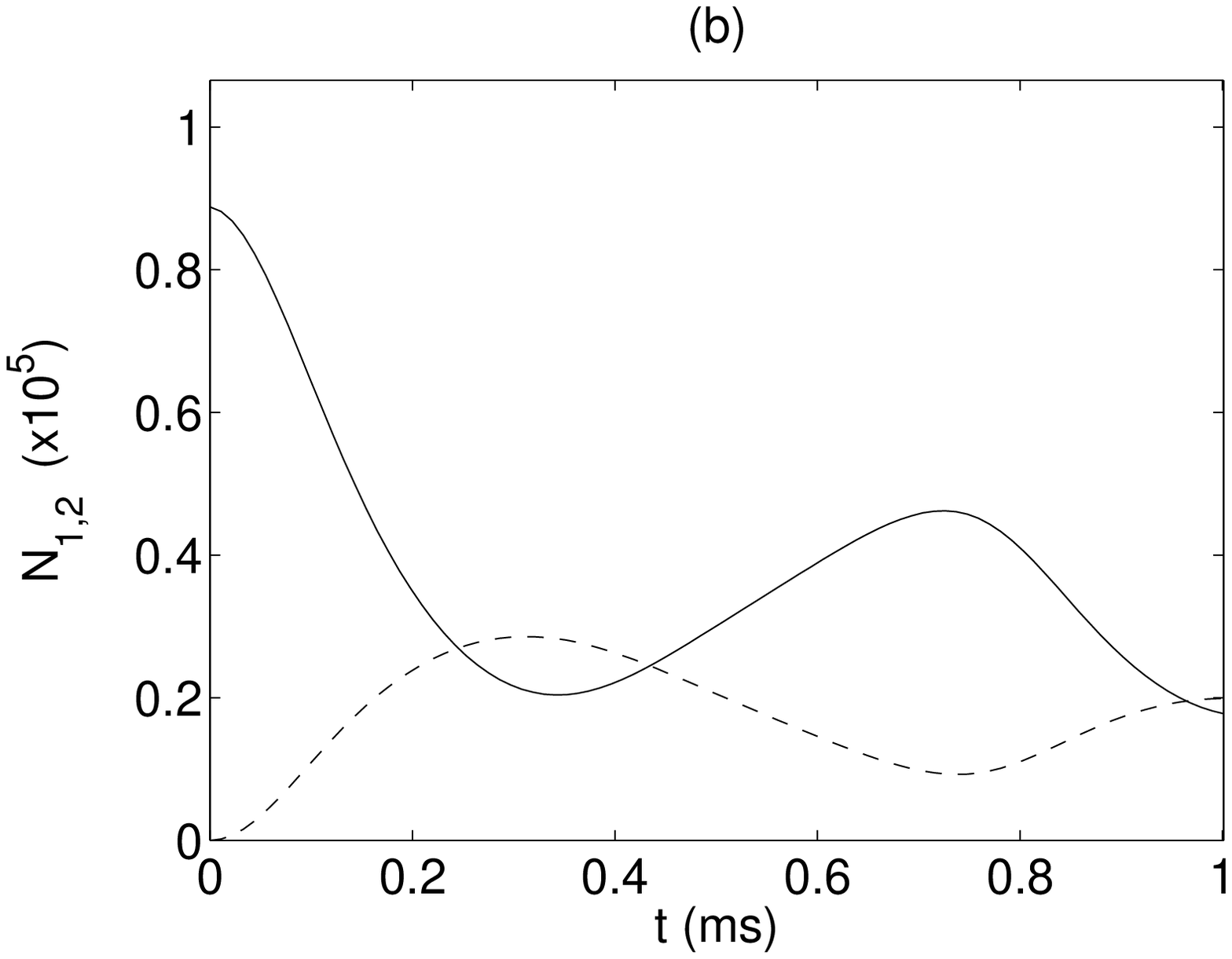,height=5 cm,width=5.7 cm}}}
\vspace{0.3 cm}
\caption{(a) Occupation numbers $N_{i}=\int d{\bf x\,}|\phi _{i}({\bf x}
,t)|^{2}$ of the atomic (solid line) and molecular (dashed line) fields, as
a function of time $t$, for the parameter values of Fig. 2; (b) same as in
(a) but for the half the initial atomic density and the same initial
effective detuning $\tilde{\delta}$.}
\end{figure}

\end{document}